%% file: main.tex
\documentclass[journal]{IEEEtran}

\ifCLASSINFOpdf
\else
   \usepackage[dvips]{graphicx}
\fi
\usepackage{url}

\usepackage{graphicx}

\usepackage{amsmath,amsfonts}
\usepackage{algorithmic}
\usepackage{algorithm}
\usepackage{array}
\usepackage{textcomp}
\usepackage{stfloats}
\usepackage{url}
\usepackage{verbatim}
\usepackage{cite}
\usepackage{textcomp}
\usepackage{xcolor}
\usepackage{tikz}
\usepackage{subcaption}

\begin{document}

\title{{Pushing the Limit of Range Resolution Beyond Bandwidth Constraint with Triangle FMCW}}

\author{
Yanbo Zhang
\thanks{Yanbo Zhang, is with Nanyang Technological University, 50 Nanyang Avenue, Singapore 639798 (e-mail: yanbo001@e.ntu.edu.sg)}
}


\maketitle

\begin{abstract}
    This paper proposes a novel signal processing technique that doubles the range resolution of FMCW~(Frequency Modulated Continuous Wave) sensing without increasing the required bandwidth. 
    The proposed design overcomes the resolution limit imposed by bandwidth by exploiting the phase consistency observed in the special frequency variation of the beat signal derived from triangle FMCW.
    Through this approach, the resolution is doubled while maintaining high spectrum SNR. 
	%
	A system model for signal processing is presented, and the energy distribution of the derived beat spectrum is analyzed. 
	The effectiveness of the proposed technique is validated through model-based simulations.
\end{abstract}


\IEEEpeerreviewmaketitle

\input{1-Introduction}

\input{2-Compare}
\input{3-Quadrature}
\input{5-Simulation_all}
\input{6-Conclusion}

\bibliographystyle{IEEEtran}
\bibliography{reference}

\end{document}

%% file: 1-Introduction.tex
\vspace{3mm}

\section{Introduction}


\IEEEPARstart{F}{MCW}~(Frequency Modulated Continuous Wave) technique has found wide applications in
wireless~sensing such as target localization~\cite{10.5555/2616448.2616478,sorrentino2012accurate,xiong2020linear}, gesture recognition~\cite{zhang2018latern,dekker2017gesture,peng2017fmcw,ryu2018feature}, fall detection~\cite{ding2019fall,peng2016fmcw,yang2018placement} and vital signal monitoring~\cite{8378778,8695699,he2017novel}. 
The performance of such systems is primarily determined by the ranging resolution of FMCW -- the capability of separating reflected signals with minimal range difference. 
To improve the resolution, conventional methods rely on the common belief between range resolution and bandwidth, by either linearly extending an FMCW symbol or increasing the linear slope of frequency variation. 
Both methods lead to increased usage of frequency bandwidth, which is however limited by available spectrum resource 
in practice. 

In this paper, we present a novel signal processing design for FMCW radar that improves the range resolution without requiring increased bandwidth.
%
%
The design is based on the observation of distinct frequency variation characteristics of the beat signal derived from triangle FMCW. 
Specifically, we find that the frequency variation of the beat signal contains distinct symmetry and continuity – in corresponding with the rising ramp and falling ramp of a triangle waveform, the frequency of the two parts of the beat signal is symmetric with respect to DC, and the frequency transition between these two parts is linear and continuous. 
These characteristics of frequency variation also translate to similar continuity and symmetry in the phase variation of the beat signal, as the signal phase is the time integral of frequency.
Further analysis indicates that these properties result in remarkable consistency of the real part of the beat signal throughout the entire symbol duration, which we refer to as \textit{phase consistency}.
Through simulations, we demonstrate that such properties are unique in beat signals generated from triangle FMCW, and are not present when using other repetitive or extended signals such as repeated upchirps (sawtooth). 
Leveraging the phase consistency, we devise an efficient signal processing method that improves the range resolution without increasing the bandwidth.
The effectiveness of our approach is demonstrated with rigorous theoretic analysis and real-world experimental validation. 

In the literature, researchers have been trying to achieve higher range resolution with given bandwidth limit by designing complex signal processing technique. 
In~\cite{li2015method}, the authors propose to utilize a similar triangle FMCW to achieve improved range resolution. However, the proposed method does not specifically handle the phase discontinuity of the complex beat signal, which makes it difficult to extract the actual beat frequency from the side lobes due to the spectrum leakage, limiting its effectiveness in multi-target scenarios. 
In~\cite{8955825}, the authors generate extended beat signal from consecutive linear chirps to improve the range resolution. 
A coherent extension scheme is proposed to avoid the discontinuity of the beat signal. 
However, such a design requires knowing the Doppler frequency shift to identify the actual multipath components, and is thus not applicable to signals reflected from static objects or those with correlated moving dymanics.  
In~\cite{4760874}, the authors propose a curve fitting based spectral analysis method to accurately identify the signal beat frequency. 
Similar ideas such as ZFFT~\cite{1170177}, FFEA~\cite{liu2005novel} and EDFT~\cite{liepins2013extended} focus on using spectral analysis for improved DFT precision. 
All such works~\cite{4760874,1170177,liu2005novel,liepins2013extended}, although with improved DFT precision, cannot fundamentally provide higher frequency resolution.  
Our proposed technique is orthogonal in its design space to these spectral enhancement approaches~\cite{4760874,1170177,liu2005novel,liepins2013extended}.

%% file: 2-Compare.tex
\vspace{3mm}

\section{Beat signal derived from different waveforms} 
\label{sec:beat signal}

\begin{figure*}[t]
	\centering
	\begin{minipage}{0.27\textwidth}
		\centerline{\includegraphics[width=\textwidth]{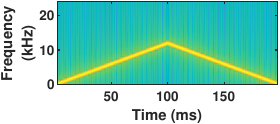}}
		\vspace{-1.5mm}
		\subcaption{}
		\vspace{-1.5mm}		
	\end{minipage}
	\begin{minipage}{0.23\textwidth}
		\centerline{\includegraphics[width=\textwidth]{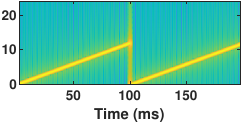}}
		\vspace{-1.5mm}
		\subcaption{}
		\vspace{-1.5mm}		
	\end{minipage}
	\begin{minipage}{0.23\textwidth}
		\centerline{\includegraphics[width=\textwidth]{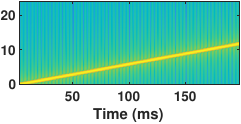}}
		\vspace{-1.5mm}
		\subcaption{}
		\vspace{-1.5mm}		
	\end{minipage}
	\begin{minipage}{0.23\textwidth}
		\centerline{\includegraphics[width=\textwidth]{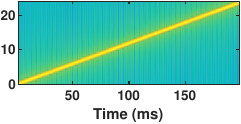}}
		\vspace{-1.5mm}
		\subcaption{}
		\vspace{-1.5mm}		
	\end{minipage}
	\vspace{-1mm}
	\caption{Spectrograms of the four FMCW variations: (a) Triangle FMCW (b) Sawtooth FMCW (c) Gentle FMCW and (d) Extended FMCW. Extended FMCW is adopted to generate a reference beat signal which can provide twice the resolution. }
	\label{f:waveforms}
\end{figure*}


\begin{figure}
	\centering
	\includegraphics[width=0.45\textwidth]{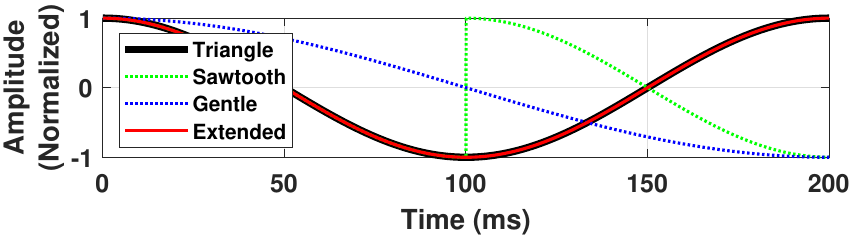}
	\caption{The time domain beat signals (real part) derived with the four types of FMCW waveforms.}
	\label{f:beat_real}
\end{figure}


The beat signal is derived by mixing the received signal with the conjugate of the transmitted signal.
The frequency components of the beat signal indicate the propagation delays of the superimposed multipath components. 
In this section, we investigate the time-domain characteristic of the beat signals derived from different FMCW waveforms.
The analysis of the characteristics is the basis for further understanding the superiority of triangle FMCW in providing improved range resolution.

We generate four variety of beat signals with using four different FMCW waveforms:
1) a rising ramp stitched with a falling ramp (triangle FMCW)
2) two repeated rising ramps (sawtooth FMCW)
3) a rising ramp with reduced slope and (gentle FMCW)
4) a rising ramp extended to twice the bandwidth (extended FMCW).
Fig.~\ref{f:waveforms} presents the spectrograms of these four FMCW variations.

We send each waveform over a fixed pre-configured wireless channel with simulation, and the received signal is mixed with the conjugate of corresponding transmitted signal to generate the beat signal.
Note that to clearly demonstrate the waveform characteristics of the different beat signals, a single-tap channel model is adopted to avoid the waveform variation caused by the superposition of different frequencies.
The channel delay is set to $\rm 1/(2B)$ for this demonstration where $\rm B$ denotes the bandwidth of triangle FMCW. 

Fig.~\ref{f:beat_real} compares the obtained beat signals (real part) in time domain.Within one symbol duration, the beat signal derived with triangle FMCW extends over a complete cycle of a single tone and aligns well with the reference beat signal. 
In contrast, the beat signal derived with sawtooth FMCW repeats its half period, and that obtained from gentle FMCW exhibits half of a doubled period. 
The different time varying patterns of these beat signals determine their different frequency resolution (although the same frequency granularity given the same amount of samples in time), which further determine the different range resolution~(as later demonstrated in Fig.~\ref{f:simu_range3}). 

The preliminary simulation demonstrates the special time-domain beat signal derived from triangle FMCW.  
In the next section, we mathematically model the beat signal and conduct a rigorous time-frequency domain analysis to explain its efficacy and the rationale behind.  

\vspace{-3mm}

%% file: 3-Quadrature.tex
\vspace{7mm}
\section{Signal Model}
\label{sec:time_resolution}


A triangle FMCW symbol contains a linear up-chirp and its subsequent mirrored down-chirp. We model the time domain waveform with the following piecewise function (assume normalized amplitude),
\begin{equation}
	x_{tri}(t)=
	\begin{cases} 
		e^{j (\pi \alpha t^2 + 2\pi f_0 t) }, & 0 < t \leq T_c \\
		e^{j [- \pi \alpha t^2 + 2\pi (f_0+B) t + \phi (T_c) ] }, & T_c < t \leq T_s
	\end{cases}
	\label{eqa:1}
\end{equation}
where $B$, $\alpha$, $f_0$ denote the bandwidth, slope and initial frequency of a single chirp. $T_s$ represents the duration of one symbol, and $T_c = T_s/2$ denotes the duration of a single chirp. As Eq.~\ref{eqa:1} indicates, triangle FMCW extends in the same way as linear FMCW before $T_c$ and then continues with reversed slope~(i.e.,~$-\alpha$) from $T_c$ to $T_s$. 
The down-chirp contains a phase delay $\phi (T_c)$ which leads to the smooth phase transition from the up-chirp. 
Without loss of generality, the formulation can be reduced as follows
\begin{equation}
	x_{tri}(t)=
	\begin{cases} 
		e^{j (\pi \alpha t^2) }, & 0 < t \leq T_c \\
		e^{j [- \pi \alpha t^2 + 2\pi B t + \pi B T_c] }, & T_c < t \leq T_s
	\end{cases}
	\label{eqa:3}
\end{equation}
where we consider single-sideband modulation (i.e., the initial frequency equals zero). The signal is transmitted and received after time delay $\tau$,~i.e., $y_{tri}(t)=x_{tri}(t-\tau)$.
Specifically,
\begin{equation}
	y_{tri}(t)=
	\begin{cases}
		0, & 0 < t \leq \tau \\
		e^{j [\pi \alpha (t-\tau)^2] }, & \tau < t \leq T_c + \tau \\
		e^{j [-\pi \alpha (t-\tau)^2+2\pi B (t-\tau) + \pi B T_c]}, & T_c + \tau < x \leq T_s
	\end{cases}
	\label{eqa:4}
\end{equation}
%
%
We derive the beat signal by mixing the received signal with the conjugate of the transmitted signal as below,
\begin{equation}
	y_{b}(t) = \overline{x_{tri}(t)} \cdot y_{tri}(t)
	\label{eqa:5}
\end{equation}
Specifically, the two signals are multiplied in each corresponding segment and thus the beat signal $y_b(t)$ is composed of three segments which are formulated as follows (the segment from 0 to $\tau$ is zero and thus is omitted here), 
\begin{equation}
	y_{b}(t)=
	\begin{cases}
		e^{j \pi (-2\alpha\tau t + \alpha {\tau}^2)}, & \tau < t \leq T_c \\
		e^{j \pi [2\alpha t^2 - ( 4B +  2\alpha \tau )t - 3B T_c + \alpha{\tau}^2 ]} , & T_c < t \leq T_c + \tau \\
		e^{j \pi (2\alpha\tau t - 4B\tau - \alpha{\tau}^2)} , & T_c+\tau < t \leq T_s
	\end{cases}
	\label{eqa:6}
\end{equation}
As we see from the above form of the produced beat signal, the frequency of $y_b(t)$ remains constant from~$\tau$ to $T_c$ and from $T_c+\tau$ to $T_s$, and linearly transitions between the two segments. Specifically, the constant frequency equals to $-k\tau$ and $k\tau$ for the first and third segment, respectively. During the second segment, the frequency increases with a slope of $2k$. 
Based on such signal model, next, we analyze the spectral leakage, explain the phase consistency of the produced beat signal, and prove the improvement of range resolution.

\vspace{0.5mm}
\noindent
\textbf{Spectral leakage.}
The beat signal is discretized and then processed by DFT to generate beat spectrum,
\vspace{-1.5mm}
\begin{equation}
    Y_{b}(k) = \mathcal{DFT}\left[y_b(n)\right] = \sum_{n=N_\tau}^{N_s} y_{b}(n) e^{\frac{-j2\pi k n}{N_s}}
    \label{eqa:7}
    \vspace{-1.5mm}
\end{equation}
where $y_b(n)$ denotes the discretized beat signal samples and $Y_b(k)$ represents the frequency component located at the $k$-th bin of the discrete spectrum, $N_s$ and $N_\tau$ denote the number~of samples that correspond to the symbol duration $T_s$ and the path delay $\tau$, respectively.
We show that the real part of the signal has a dominating frequency of $\alpha \tau$ and with little spectral leakage. 

Assume $T_c \gg \tau$ (i.e., $N_c \gg N_\tau$, the assumption can be satisified for most wireless sensing applications~\cite{enwiki:1027754964}), 
the second segment of $y_b(n)$ is ignorable and Eq.~\ref{eqa:7} can be approximated as, 
\begin{equation}
	Y_{b}(k) = \sum_{n=N_\tau}^{N_c} y_{b}(n) e^{\frac{-j2\pi k n}{N_s}} + \sum_{n=N_c+N_\tau}^{N_s} y_{b}(n) e^{\frac{-j2\pi k n}{N_s}}
	\label{eqa:8}
\end{equation}
Let $Y_{\Re_b}(k)$ be the DFT of real part of the beat signal.~According to the property of DFT, we can formulate~$Y_{\Re_b}(k)$ as,
\begin{equation}
	Y_{\Re_b}(k) = \dfrac{1}{2} [Y_{b}(k) + \overline{Y_{b}(-k)}]
    \label{eqa:extra}
\end{equation}
where $\overline{Y_{b}(-k)}$ denotes the conjugate of the beat spectrum at the negative frequency. 
By incorporating Eq.~\ref{eqa:8} to Eq.~\ref{eqa:extra}, the frequency component at $k=\alpha \tau$ can be approximated as

\vspace{-3mm}
\begin{equation}
	Y_{\Re_b}(\alpha \tau) = (N_c-N_\tau)e^{-j\pi \alpha\tau^2}
	\label{eqa:16}
\end{equation}
When $N_c \gg N_\tau$, the energy of the frequency component at $k = \alpha \tau$ can be calculated as below,
\begin{equation}
	E_{Y_{\Re_{b}}(\alpha \tau)} = (|Y_{\Re_b}(\alpha \tau)|+|Y_{\Re_b}(-\alpha \tau)|)^2 \approx |N_s|^2
	\label{eqa:17}
\end{equation}
Eq.~\ref{eqa:17} indicates that the energy of the frequency component at $k = \alpha \tau$ greatly dominates the frequency spectrum~(almost 100\%) and thus spectral leakage is ignorable. 
Next, we explain the underlying rationale of such low spectral leakage. 

\vspace{0.5mm}
\noindent
\textbf{Phase consistency.}
Despite the separation of the frequency varying segment (the impact of this segment is analyzed in Sect.~\ref{sec:simulation_all}), the phase of the real part of the third segment well aligns with the phase that is continuously extended from the first segment. 
We name this observation as phase consistency.
Because of such a special characteristic, the real part of the derived beat signal extends over a full period within the DFT window (as Fig.~\ref{f:beat_real} already suggests), which takes effect to avoid the spectral leakage.

To prove the phase consistency, we derive the initial phase of the third segment (i.e., $t = T_c+\tau$) with Eq.~\ref{eqa:6} as below,
\begin{equation}
	\phi_{y_b(T_c+\tau)} = \angle \left[ e^{j \pi (2 \alpha \tau t - 4B\tau - \alpha{\tau}^2)} \right]_{t=T_c+\tau} = \pi (k {\tau}^2-2B\tau)
	\label{eqa:18_1}
\end{equation}
Assuming the beat signal of the first segment extends continuously to the third segment~(like the case of using linear FMCW of twice the bandwidth), we derive the phase at time~$T_c+\tau$,
\begin{equation}
	{\phi}^{'}_{y_b(T_c+\tau)} = \angle \left[ e^{j \pi (-2k\tau t + k {\tau}^2)} \right]_{t=T_c+\tau} = \pi (-k {\tau}^2-2B\tau)
	\label{eqa:19}
\end{equation}
When $\tau=p/(2B), p \in \mathbb{Z^+}$, 
\vspace{-2mm}
\begin{equation}
		\phi_{y_b(T_c+\tau)} = -{\phi}^{'}_{y_b(T_c+\tau)}
	\label{eqa:20}
\end{equation}
%
%
Eq.~\ref{eqa:20} indicates that the real part (cosine) of the third segment maintains consistent phase with that of the first segment. 
Therefore, the two segments connect smoothly and perform as the natural time-domain extension, which directly results in the improved resolution on frequency domain (as later proved). 
We demonstrate the improvement in Sect.~\ref{sec:simulation_all} where we also analyze the impact when $p \notin \mathbb{Z^+}$. 

\begin{figure}[t]
    \vspace{-1mm}
	\centering
     \includegraphics[width=0.485\textwidth]{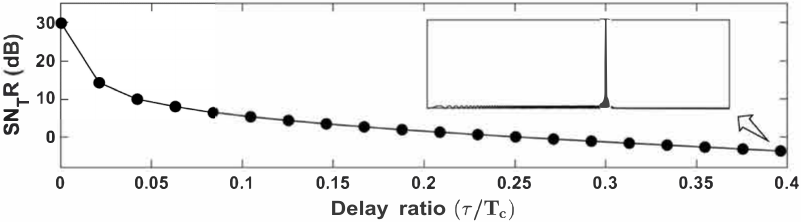}
     \caption{The impact of increasing propagation delay on $\rm SN_{T}R$. }
	\label{f:spectrum_SNR}
\end{figure}

\vspace{0.5mm}
\noindent
\textbf{Range resolution.}
The above analysis leads to direct proof in the improvement of range resolution. 

By performing DFT on the beat signal, we separate the frequency components contained in the beat signal, which map to different multipath ranges by multiplying a scaling factor. 
Therefore, the range resolution~($\Delta R_{min}$) is fundamentally determined by the frequency resolution~($\Delta f_{min}$) of the beat signal, specifically, 
\vspace{-1mm}
\begin{equation}
    \Delta R_{min} = \Delta f_{min} \cdot c / \alpha
    \label{eqa:range_freq}
    \vspace{-1mm}
\end{equation}
where $c$ denotes the signal propagation speed.
In the equation, the frequency resolution $\Delta f_{min}$ represents the ability to distinguish two closely spaced frequency components of the original signal, which equals the inverse of the measurement duration of each frequency beat ($T_{b}$), i.e., $\Delta f_{min} = 1/T_{b}$. 
Note that, $T_b$ represents the actual measurement duration that a time-domain signal is continuously sampled over, which may not equal the length of DFT window (which instead determines the granularity of frequency domain signal after DFT). 

With phase consistency, the real part of the beat signal effectively emulates a set of frequency beats which are continuously sampled over twice the chirp duration and thus $T_b=2T_c$. Therefore, the range resolution can be further derived as below,
\vspace{-2mm}
\begin{equation}
    \Delta R_{min} = 1/T_b \cdot c / \alpha = 1/T_b \cdot c \cdot T_c/B = c/2B
    \label{eqa:range_resolution}
\end{equation}
which is improved from $c/B$~(provided by the standard method) to $c/2B$ by two times.

\vspace{-2mm}

%% file: 5-Simulation_all.tex
\vspace{5mm}
\section{Simulation}
\label{sec:simulation_all}
In this section, we adopt simulation to demonstrate the effectiveness of the proposed method by comparing its ranging resolution with those obtained from other approaches as listed below: 
1) using the same triangle FMCW but adopting a different processing method~\cite{li2015method} (referred to as triangle FMCW-Li)
2)~adopting our proposed processing method but using sawtooth FMCW
and 3) adopting our method but using gentle FMCW.
The spectrograms of sawtooth and gentle FMCW are as presented in Fig.~\ref{f:waveforms}.

We send the variations of FMCW waveforms over a four-tap Rayleigh channel where the multipath profile 
can be self defined. 
The received signals 
are processed with the four methods to derive separate range profiles which depict the distribution of multipath components at different ranges. 

\begin{figure*}[t]
	\centering
	\begin{minipage}{0.26\textwidth}
		\centerline{\includegraphics[width=\textwidth]{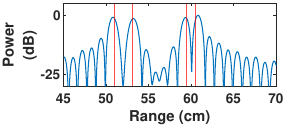}}
		\vspace{-1mm}
		\subcaption{triangle FMCW}
		\vspace{-1.5mm}		
	\end{minipage}
	\begin{minipage}{0.23\textwidth}
		\centerline{\includegraphics[width=\textwidth]{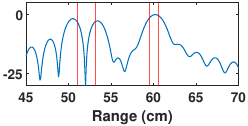}}
		\vspace{-1mm}
		\subcaption{triangle FMCW-Li}
		\vspace{-1.5mm}		
	\end{minipage}
	\begin{minipage}{0.23\textwidth}
		\centerline{\includegraphics[width=\textwidth]{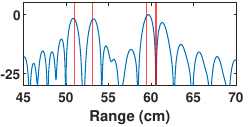}}
		\vspace{-1mm}
		\subcaption{sawtooth FMCW}
		\vspace{-1.5mm}		
	\end{minipage}
	\begin{minipage}{0.23\textwidth}
		\centerline{\includegraphics[width=\textwidth]{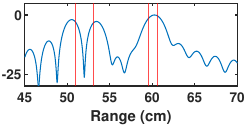}}
		\vspace{-1mm}
		\subcaption{gentle FMCW}
		\vspace{-1.5mm}		
	\end{minipage}
	\caption{The range profiles which depict the distribution of multipath at different ranges that are obtained with the four methods. 
    The ground-truth ranges are annotated with red lines.
    Using triangle FMCW of our design can separate the four paths with the highest resolution. }
	\label{f:simu_range3}
\end{figure*}

Fig.~\ref{f:simu_range3} demonstrates the range profiles obtained with the four different FMCW waveforms. 
With the profile obtained from triangle FMCW (Fig.~\ref{f:simu_range3}a), we can clearly distinguish the four paths at the ranges of 51cm, 53cm, 60cm, and 61cm (with $p$ equalling 48, 50, 56, 57 in respective). The path range estimations are well aligned with the ground-truth (annotated with the vertical red lines). 
Fig.~\ref{f:simu_range3}b presents the results obtained with triangle FMCW-Li, which fails to reflect the correct number of multipath components, and cannot provide sufficient range resolution to resolve the third and fourth path (the two paths merge into one peak). 
Fig.~\ref{f:simu_range3}c depicts the profile obtained with sawtooth FMCW. 
Similarly it fails to resolve the third and forth paths when they are close (1cm range difference).
Fig.~\ref{f:simu_range3}d shows the profile obtained with gentle FMCW. The granularity of this profile is coarser (half range granularity) because the slope of gentle FMCW is half of the other waveforms. As a result its resolution of this result cannot exceed 2cm. 
Overall, the range profile obtained with the triangle FMCW provides twice the range resolution compared to the other results. 

\vspace{-1mm}

\begin{figure*}[t]
	\centering
	\begin{minipage}{0.25\textwidth}
		\centerline{\includegraphics[width=\textwidth]{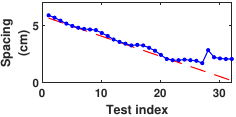}}
		\vspace{-1mm}
		\subcaption{extended FMCW (oracle)}
		\vspace{-1.5mm}		
	\end{minipage}
	\begin{minipage}{0.24\textwidth}
		\centerline{\includegraphics[width=\textwidth]{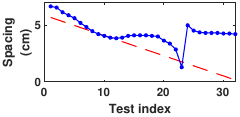}}
		\vspace{-1mm}
		\subcaption{linear FMCW (baseline)}
		\vspace{-1.5mm}		
	\end{minipage}
	\begin{minipage}{0.24\textwidth}
		\centerline{\includegraphics[width=\textwidth]{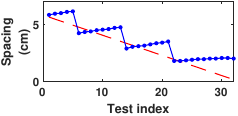}}
		\vspace{-1mm}
		\subcaption{triangle FMCW}
		\vspace{-1.5mm}		
	\end{minipage}
	\begin{minipage}{0.23\textwidth}
		\centerline{\includegraphics[width=\textwidth]{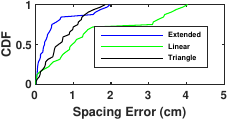}}
		\vspace{-1mm}
		\subcaption{CDF of the spacing error}
		\vspace{-1.5mm}		
	\end{minipage}
	\caption{Estimation of the spacing between two close targets. The spacing is gradually reduced over the 30 tests. (a) to (c) detail the results obtained with different FMCW waveforms (ground truth is annotated in red dashed line) and (d) illustrates their statistical spacing errors. }
	\vspace{-4mm}
	\label{f:simu_spacing}
\end{figure*}

\begin{figure}[t]
    \vspace{-1mm}
	\centering
     \includegraphics[width=0.485\textwidth]{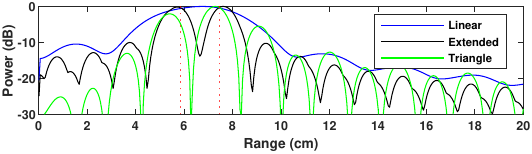}
     \vspace{-2mm}
     \caption{Range profile derived with non-integer $p$. } 
	\label{f:non-integer profile}
\end{figure}

\vspace{0.5mm}
\noindent
\textbf{Impact of non-integer $p$.}
If $p \notin \mathbb{Z^+}$, the fractional residual causes mismatched phase at the junction of the two segments. 
To evaluate the impact of this mismatch, we set the range of the two paths to 5.9cm and 8.2cm in our simulation, and then performed multipath separation by using three different types of signals: linear FMCW, extended FMCW and triangle FMCW. 
The results of multipath separation is shown in Fig.~\ref{f:non-integer profile}. 
As we see from the results, despite the deviation in accuracy -- which is due to the loss of phase consistency when $p$ is non-integer, the results derived with both extended FMCW and triangle FMCW identifies the two peaks successfully. 
Therefore, the improved resolution of extended FMCW remains even when $p$ is a non-integer value. 
However, it is equally important to note that in this case, the accuracy of the range estimation may get reduced.

\vspace{0.5mm}
\noindent
\textbf{Impact of increased target range.}
The presence of the frequency varying segment raises~the noise floor of the beat spectrum and the impact may intensify as target range increases. 
We conduct a further simulation to assess how the increasing range impacts on the spectrum $\rm SN_{T}R$~($\rm N_T$ denotes the spectrum noise induced by the transition period). 
Specifically, a triangle FMCW signal is sent over a single-tap channel model where the path delay increases. 
We record the $\rm SN_{T}R$ of each measurement and the result is depicted in Fig.~\ref{f:spectrum_SNR}. 
Although $\rm SN_{T}R$ reduces with increasing range, 
it is still over -7.4~dB~(the spike can still be detected at this $\rm SN_{T}R$) when the delay exceeds 40\% of the chirp duration,
which equivalently supports km-level target range for radio frequency based sensing applications. 

%% file: 6-Conclusion.tex
\section{Conclusion}
\label{sec:conclusion}

In this paper, we propose a novel signal processing technique that 
leverages the phase consistency of beat signal derived with triangle FMCW to improve the resolution of FMCW ranging with signal bandwidth limit. We prove the efficacy of the design with spectral analysis and evaluate the improvement with model based simulation and real world experiment.
The range accuracy of the proposed technique may reduce when the actual range does equal integer number of the resolution. This issue can be mitigated by applying calibration or filtering algorithms to correct the estimation results. 
The accuracy could also be affected by the Doppler effect, which may be resolved by incorporating the standard triangle FMCW based Doppler estimation approach. 
We leave the detailed study on this issue as future work.